\documentclass[aps,prx,reprint,superscriptaddress]{revtex4-1}

\usepackage{graphicx}
\usepackage{dcolumn}
\usepackage{amsmath, amscd, amssymb, amsthm, bm}

\usepackage[dvipsnames]{xcolor}

\usepackage{microtype}  
              
\makeatletter
\def\textless{\afterassignment\textless@\let\next= }
\def\textless@#1#{\@nameuse{textless@#1}}
\def\textless@sub#1#2/sub#3{%
  \ensuremath{_{\let\textless\relax#2}}%
  \egroup 
}
\def\textless@sup#1#2/sup#3{%
  \ensuremath{^{\let\textless\relax#2}}%
  \egroup 
}
\makeatother

\begin{document}


\title{Influence of spin and orbital fluctuations on Mott-Hubbard exciton dynamics in LaVO${}_{3}$ Thin Films}

\author{D. J. Lovinger}
\affiliation{Department of Physics, University of California, San Diego, La Jolla, California 92093}
\author{M. Brahlek}
\affiliation{Department of Materials Science and Engineering, The Pennsylvania State University, University Park, Pennsylvania 16801}
\author{P. Kissin}
\affiliation{Department of Physics, University of California, San Diego, La Jolla, California 92093}
\author{D. M. Kennes}
\affiliation{Institut f\"{u}r Theorie der Statistischen Physik, RWTH Aachen University
and JARA-Fundamentals of Future Information Technology, Aachen, Germany 52056}
\affiliation{Max Planck Institute for the Structure and Dynamics of Matter, Center
for Free Electron Laser Science, Hamburg, Germany 22761}
\author{A. J. Millis}
\affiliation{Department of Physics, Columbia University, New York, New York 10027}
\author{R. Engel-Herbert}
\affiliation{Department of Materials Science and Engineering, The Pennsylvania State University, University Park, Pennsylvania 16801}
\affiliation{Department of Physics, The Pennsylvania State University, University Park, Pennsylvania 16801}
\affiliation{Department of Chemistry, The Pennsylvania State University, University Park, Pennsylvania 16801}
\author{R. D. Averitt}
\affiliation{Department of Physics, University of California, San Diego, La Jolla, California 92093}

\date{\today}

\begin{abstract}
Recent optical conductivity measurements reveal the presence of Hubbard excitons in certain Mott insulators. In light of these results, it is important to revisit the dynamics of these materials to account for excitonic correlations. We investigate time-resolved excitation and relaxation dynamics as a function of temperature in perovskite-type LaVO${}_{3}$ thin films using ultrafast optical pump-probe spectroscopy. LaVO${}_{3}$ undergoes a series of phase transitions at roughly the same critical temperature $T_C\cong 140\ K$, including a second-order magnetic phase transition (PM $\xrightarrow{}$ AFM) and a first-order structural phase transition, accompanied by \textit{C}-type spin order (SO) and \textit{G}-type orbital order (OO). Ultrafast optical pump-probe spectroscopy at 1.6 eV monitors changes in the spectral weight of the Hubbard exciton resonance which serves as a sensitive reporter of spin and orbital fluctuation dynamics. We observe dramatic slowing down of the spin, and orbital dynamics in the vicinity of $T_C\cong 140$ K, reminiscent of a second-order phase transition, despite the (weakly) first-order nature of the transition. We emphasize that since it is spectral weight changes that are probed, the measured dynamics are not reflective of conventional exciton generation and recombination, but are related to the dynamics of Hubbard exciton formation in the presence of a fluctuating many-body environment.
\end{abstract}

\maketitle

\section{Introduction}
\label{sec:LVOintroduction}

Transition-metal oxides (TMO) display a wide range of interesting magnetic and electronic properties, owing to strong electron correlation and competition between charge, spin, lattice, and orbital degrees of freedom. The interplay of these interactions makes for rich and diverse physics, with relevant phenomena including colossal magnetoresistance, metal-insulator transitions, ferroelectricity, and high-\textit{T${}_{c}$} superconductivity \cite{Basov2011,Imada1998}. The rare-earth vanadates RVO${}_{3}$ (R = rare-earth or Y) are a class of TMO displaying many of these features. Specifically, these materials are Mott-Hubbard (MH) insulators in which spin, charge, and orbital degrees of freedom are intimately coupled, making them an ideal prototypical system for studying the interplay of some of the most fundamental properties of matter \cite{Raychaudhury2007,Park2017,Jackeli2008,Varignon2015, Brahlek2017}. 

Recently, optical conductivity measurements and DFT/DMFT calculations indicate that Hubbard excitons (HE) play an important role in the physics of vanadates \cite{Novelli2012a,Reul2012a,Kim2018a}. The lowest lying peak in $\sigma_{1}$ ($\sim$1.8 eV), previously attributed to a 3\textit{d} multiplet, appears to belong instead to an excitonic resonance. Further, the formation of these excitonic signatures appear to be strongly influenced by the presence of spin and orbital order. In general, there remain significant open questions regarding the physics of MH excitons in quantum materials. Under what conditions do these excitons form, what are their characteristic timescales, and what degrees of freedom influence their dynamics? Advances in theory and experimental techniques have begun to address these questions, indicating for example that the recombination rates are proportional to the MH gap and mediated by magnon emission \cite{Lenarcic2013,Prelovsek2016,Wrobel2002,Li2018}. Ultrafast optical spectroscopy is a technique particularly well-suited to furthering our understanding of this field, and has been used extensively to study ultrafast dynamics in many MH insulators, including cuprates \cite{Sahota2019,Schneider2002,Vishik2017}, manganites \cite{Lobad2000c,Zhang2016}, and TaS$_{2}$ \cite{Mann2016a}.

\begin{figure*}
  \centering
  \includegraphics[width=0.9\textwidth]{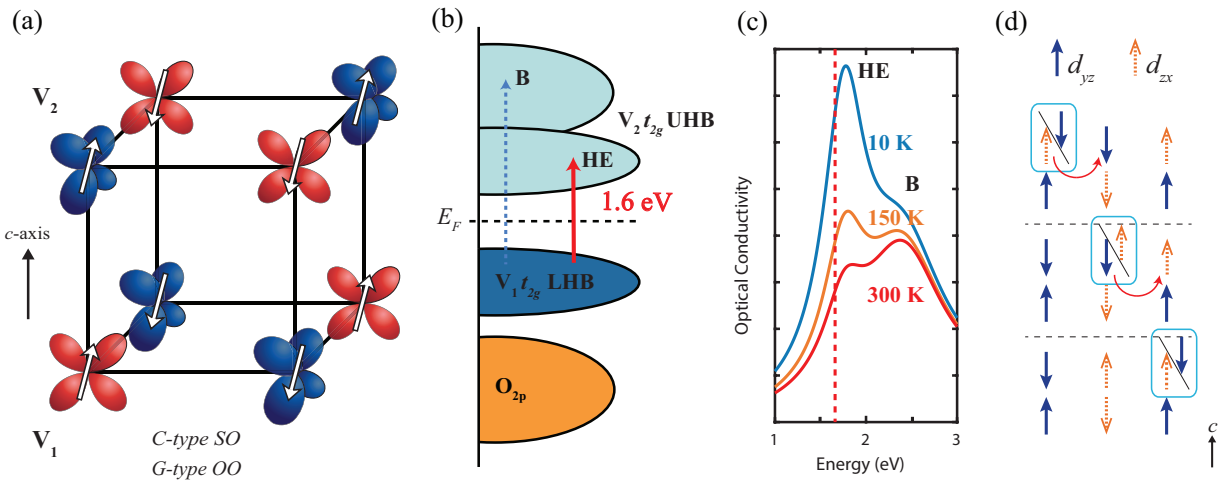}
  \caption[(a) Diagram of the C-type spin and G-type orbital order configuration in LaVO$_3$.]
  {(a) Diagram of the C-type spin and G-type orbital order configuration in LaVO$_3$. The spins on $d_{zx}$ and $d_{yz}$ orbitals, shown as white arrows, are slightly canted from the \textit{c}-axis. (b) Depiction of the density of states near the Fermi level, showing the photoexcitation process at 1.6 eV from the lower Hubbard band (LHB) to the upper Hubbard band (UHB) as a solid red arrow, corresponding to peak HE in the optical conductivity. An electron from a $t_{2g}$ orbital in the LHB on Vanadium atom V$_1$ is promoted to a $t_{2g}$ orbital in the UHB on V$_2$. The dashed arrow depicts the single particle excitation with well-separated quasiparticles, corresponding to peak B in the optical conductivity. (c) Depiction of the \textit{c}-axis optical conductivity evolution with temperature, reproduced from \cite{Miyasaka2002}. Peak HE corresponds to the excitonic resonance in the vicinity of the MH gap, and peak B an excitation from LHB $\xrightarrow{}$ UHB (consisting of well separated quasiparticles). The ratio of peak HE to peak B grows drastically below the ordering temperature at $T_C$ = 140 K, indicating the importance of spin and orbital order upon the exciton spectral weight. The red dotted line indicates the laser photoexcitation energy. (d) Motion of a ``double occupancy" on the spin and orbitally ordered lattice. Each hopping process creates a trace of disordered spins and orbitals \index{LVO1_OO}}
  \label{fig:LVO1_OO}
\end{figure*}

The relative simplicity of the vanadates as \textit{d${}^{2}$} materials, along with coupled spin and orbital order, makes them an excellent platform in which to study HE dynamics, particularly in relation to the spin and orbital degrees of freedom. In this work we focus on lanthanum vanadate LaVO${}_{3}$ (LVO). LVO has a perovskite-type lattice and is a \textit{3d${}^{2}$} Mott-Hubbard insulator, with both d-electrons occupying the \textit{t${}_{2g}$} band. Strong thermal and quantum orbital fluctuations, enhanced by a highly frustrated Kugel-Khomskii superexchange \cite{Weng2010a}, play a role in stabilizing the \textit{t${}_{2g}$} energy even at room temperature, resulting in fairly equal occupation of the \textit{d${}_{xy}$}, \textit{d${}_{yz}$}, and \textit{d${}_{zx}$} orbitals \cite{Horsch2008,Khaliullin2001,Raychaudhury2007}. Upon cooling, this degeneracy is broken as LVO undergoes a structural phase transition at \textit{T${}_{C}$} = 140 K that, in turn, modifies the spin and orbital order. A first-order structural phase transition from an orthorhombic \textit{Pbnm }lattice to monoclinic P2${}_{1}$/b \cite{Bordet1993,Miyasaka2002} is accompanied by G-type Jahn-Teller and GdFeO${}_{3}$-type distortion. As shown in Fig. \ref{fig:LVO1_OO}(a), this lifts the \textit{t${}_{2g}$} orbital degeneracy and results in G-type orbital ordering with an always occupied \textit{d${}_{xy}$} orbital and alternatingly occupied \textit{d${}_{yz}$} and \textit{d${}_{zx}$} orbitals in all lattice directions \cite{Khaliullin2001,Miyasaka2002,Sawada1996,Kim2018a}. This orbital configuration subsequently induces an antiferromagnetic exchange interaction between sites in the \textit{ab} plane, resulting in a transition which has been theoretically argued to be second-order from paramagnetic to C-type antiferromagnetic order with slightly canted spins \cite{Khaliullin2001}.

The degree of spin and orbital order also affects the optical properties of LVO, which exhibits a strong anisotropy in the optical conductivity. While the spectrum is almost entirely temperature independent for $\boldsymbol{E}\bot c$, for $\boldsymbol{E}\parallel c$ there is a large transfer of spectral weight from high ($\mathrm{>}$ 3.5 eV) to low ($\mathrm{\sim}$2 eV) energy as the temperature is lowered, as shown in Fig. \ref{fig:LVO1_OO}(c) (data reproduced from  \cite{Miyasaka2002}). Furthermore, a splitting of the 2 eV region is observed, separating the low energy optical conductivity into peak HE at 1.8 eV and peak B at 2.4 eV. Both correspond to an excitation across the MH gap, associated with \textit{d${}_{yz}$}-\textit{d${}_{yz}$} or \textit{d${}_{zx}$}-\textit{d${}_{zx}$} transitions between adjacent V${}^{3+\ }$sites along the \textit{c}-axis \cite{Miyasaka2002,Tomimoto2003}, resulting in a high-spin excited state. While peak B represents a single particle excitation consisting of a well separated double occupancy and holon, peak HE reflects the Hubbard exciton, with the peak separation proportional to the HE binding energy \cite{Reul2012a,Kim2018a}. This distinction is clarified in Fig. \ref{fig:LVO1_OO}(b), with the 1.6 eV red arrow indicating our experimental pump-probe energy at the HE peak (as described in greater detail below). As peak HE lies above the MH gap \cite{Arima1993a}, it is not a truly bound state. Rather, it is an excitonic resonance within the continuum, comprised of a weakly bound state between an excited \textit{d${}^{3}$} state in the upper Hubbard band and a \textit{d${}^{1}$} state in the lower Hubbard band \cite{Novelli2012a,Reul2012a}. 

It is of particular importance to note the influence of spin and orbital order upon Hubbard exciton dynamics. In the following, we refer to the photoexcited quasiparticle as a ``double occupancy'' (DO), but note that it is composed of two half-filled \textit{d${}_{yz}$} and \textit{d${}_{zx}$} orbitals, each with a single electron, and not a traditional doublon. Exciton formation in semiconductors is generally driven by a lowering of the Coulomb energy, but in an antiferromagnetic Mott-Hubbard system it is governed by the kinetic energy. Hopping of a single holon or DO on the AFM background disrupts the local spin order, leaving a trace of disordered spins behind [Fig. \ref{fig:LVO1_OO}(d)]. As each step requires more energy than the last due to the compounding magnetic frustration, motion of a bare DO or holon is hindered and the kinetic energy reduced. However, moving as a pair preserves spin order, so a spinless bound exciton may freely move through the lattice, thereby gaining kinetic energy and becoming the energetically favorable state \cite{Wrobel2002,Clarke1993a}. A similar argument may be made for the motion of holons in an antiferro-orbitally ordered background (as in LVO) where motion of the holon or DO leaves a trace of misaligned orbitals \cite{Gossling2008a}. Indeed, static optical conductivity measurements have shown the influence of orbital order in Hubbard exciton formation in vanadates such as YVO${}_{3}$, GdVO${}_{3}$, and CeVO$_{3}$ \cite{Reul2012a}. In addition, time-resolved optical conductivity measurements on YVO${}_{3}$ indicate the role of spin order/disorder on exciton dynamics \cite{Novelli2012a}. 

It follows that optical measurements tuned near the excitonic peak at 1.8 eV serve as a highly sensitive probe of the coupled spin and orbital order. In the present work we carry out measurements using femtosecond optical pump probe spectroscopy in the vicinity of the HE resonance in epitaxial LaVO$_{3}$ thin films grown on SrTiO$_{3}$ to monitor spin and orbital dynamics. The various processes involved in excitation and recovery have unique intrinsic timescales, including spin-lattice coupling on the order of $\mathrm{<}$ 10 ps and recovery of spin/orbital order in $\mathrm{>}$ 500 ps, allowing us to disentangle interacting degrees of freedom and observe how the dynamics are altered in different phases. In doing so we observe an anomalous slowdown in the dynamics near the spin and orbital ordering phase transition, attributable in part to the influence of a highly fluctuating spin/orbital background on the presence of Hubbard excitons.

\section{Methods}
\label{sec:LVOmethods}

The LaVO${}_{3}$ sample studied is a $\mathrm{\sim}$50 nm thin film grown by hybrid molecular beam epitaxy on a (001) SrTiO${}_{3}$ substrate \cite{Zhang2015,Brahlek2018,Zhang2017}. X-ray diffraction measurements have confirmed that the LVO film is coherently strained to the substrate. The film is compressively strained, with an out of plane lattice parameter of 3.951 {\AA} and in-plane lattice parameter of 3.905 {\AA}, yielding an epitaxial strain of \textit{c/a} = 1.012. As the film is grown on a higher symmetry substrate, LVO forms rotational domains with the crystallographic '+' axis, or \textit{c}-axis, in plane, oriented orthogonal to one another. Since the \textit{c}-axis is in-plane the pump and probe beam always have a projection along \textit{c}, and are therefore sensitive to the spectral weight changes. Consistent with rotational domains, the signal does not change with rotation of the sample. 

\begin{figure}[t] 
  \centering
  \includegraphics[width=\linewidth]{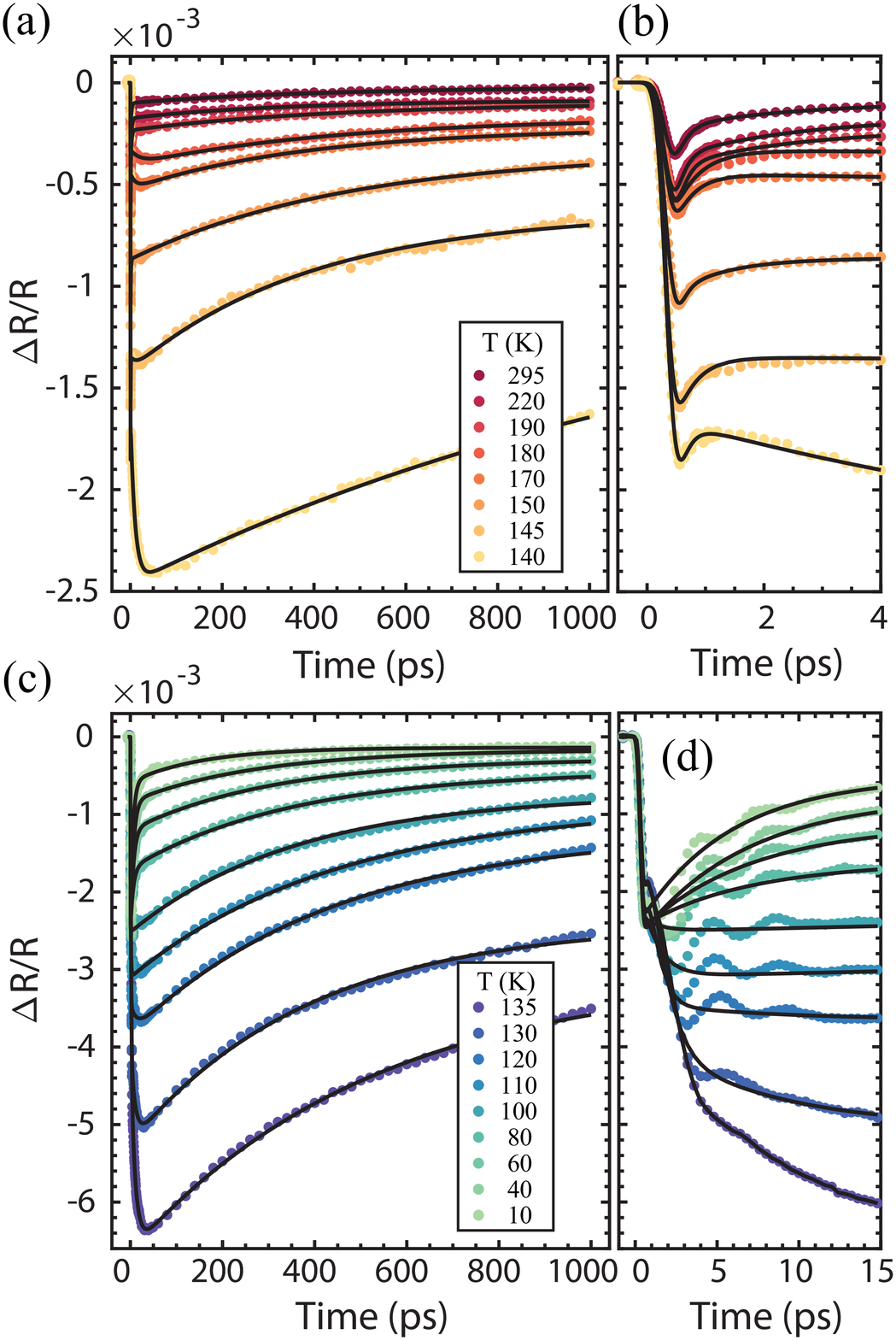}
  \caption[Photoinduced reflectivity traces in the high temperature phase (T $\geq$ 140 K), for long (a) and short (b) pump-probe delays, and in the low temperature phase (T $\leq$ 135 K) for long (c) and short (d) pump-probe delays.]
  {Photoinduced differential reflectivity traces in the high temperature phase (T $\geq$ 140 K), for long (a) and short (b) pump-probe delays, and in the low temperature phase (T $\leq$ 135 K) for long (c) and short (d) pump-probe delays. The black curves are exponential fits to the data, of the form given in Eq. \ref{eq:LVO1_dRR}. \index{LVO2_drrAll}}
  \label{fig:LVO2_drrAll}
\end{figure}

The ultrafast optical measurements were made using a Spectra-Physics Spirit 200 kHz 1040 nm Yb-based hybrid-fiber laser coupled to a nonlinear optical parametric amplifier. The amplifier outputs $\mathrm{\sim}$20 fs pulses at 770 nm (1.61 eV), which are split, cross-polarized (pump \textit{s-}polarized, probe \textit{p}), and used as degenerate pump and probe beams in a reflective geometry. We used a low pump fluence of 60 $\mu$J/cm${}^{2}$ to ensure that we were in the linear excitation regime. Measurements at a fluence of 20 $\mu$J/cm${}^{2}$ show no change in the dynamics in comparison to the 60 $\mu$J/cm${}^{2}$, indicating that the laser heating is small. The laser energy of 1.61 eV is near the HE peak in the optical conductivity, as discussed above [Fig. \ref{fig:LVO1_OO}(c)]. Photoexcitation at this wavelength corresponds to an inter-site V${}^{3+}$ \textit{d-d} transition.  In the orbitally ordered state this involves electron transfer between \textit{d${}_{yz}$-d${}_{yz}$} or \textit{d${}_{zx}$-d${}_{zx}$} orbitals on adjacent vanadium sites along the \textit{c}-axis \cite{Sawada1996}, as shown in Fig. \ref{fig:LVO1_OO}(b). The formation of Hubbard excitons is highly sensitive to spin and orbital order (i.e. the spectral weight of the HE peak in Fig. \ref{fig:LVO1_OO}(c) decreases with increasing temperature because of spin and orbital fluctuations). Therefore, by measuring changes in the reflectivity after photoexcitation as a function of pump-probe delay time (arising from photoinduced HE spectral weight transfer), we can track the time-dependent dynamics associated with spin, orbital, and structural order in LVO at various temperatures.

\section{Experimental Results}
\label{sec:LVOResults}

The time-dependent differential reflectivity signal $\Delta R/R$ in the high temperature phase (T $\mathrm{\ge}$ 140 K) is shown in Fig. \ref{fig:LVO2_drrAll}(a) and (b), for long and short pump-probe delays, respectively. The black lines are exponential fits to the data. Generally, the photoinduced change in $\Delta R/R$ is negative; there is a sharp increase in the signal amplitude occurring in $\mathrm{\sim}$500 fs, followed by a slower multi-component recovery. As the temperature is lowered from 295 K we observe nearly an order of magnitude increase in the signal amplitude and a slowing of the recovery dynamics. Remarkably, upon crossing \textit{T${}_{C}$} = 140 K the trend is reversed. As shown in Fig. \ref{fig:LVO2_drrAll}(c) and (d), the signal amplitude decreases and the dynamics accelerate as the sample is cooled to 10 K. This indicates a fundamental change in the degrees of freedom our probe is sensitive to upon traversing \textit{T${}_{C}$}. Also of note is the emergence of an overdamped coherent mode with a period of $\mathrm{\sim}$4 ps, seen in Fig. \ref{fig:LVO2_drrAll}(d), which becomes more pronounced as the temperature is lowered. This is the result of an acoustic phonon launched by photoexcitation at the sample surface \cite{C.ThomsenH.T.GrahnH.J.Maris1986a}. The generation and propagation of the coherent acoustic phonon is affected by the phase transitions occurring at $T_C$, but this is beyond the scope of the present manuscript and will not be further discussed. 

\begin{figure}[t] 
  \centering
  \includegraphics[width=\linewidth]{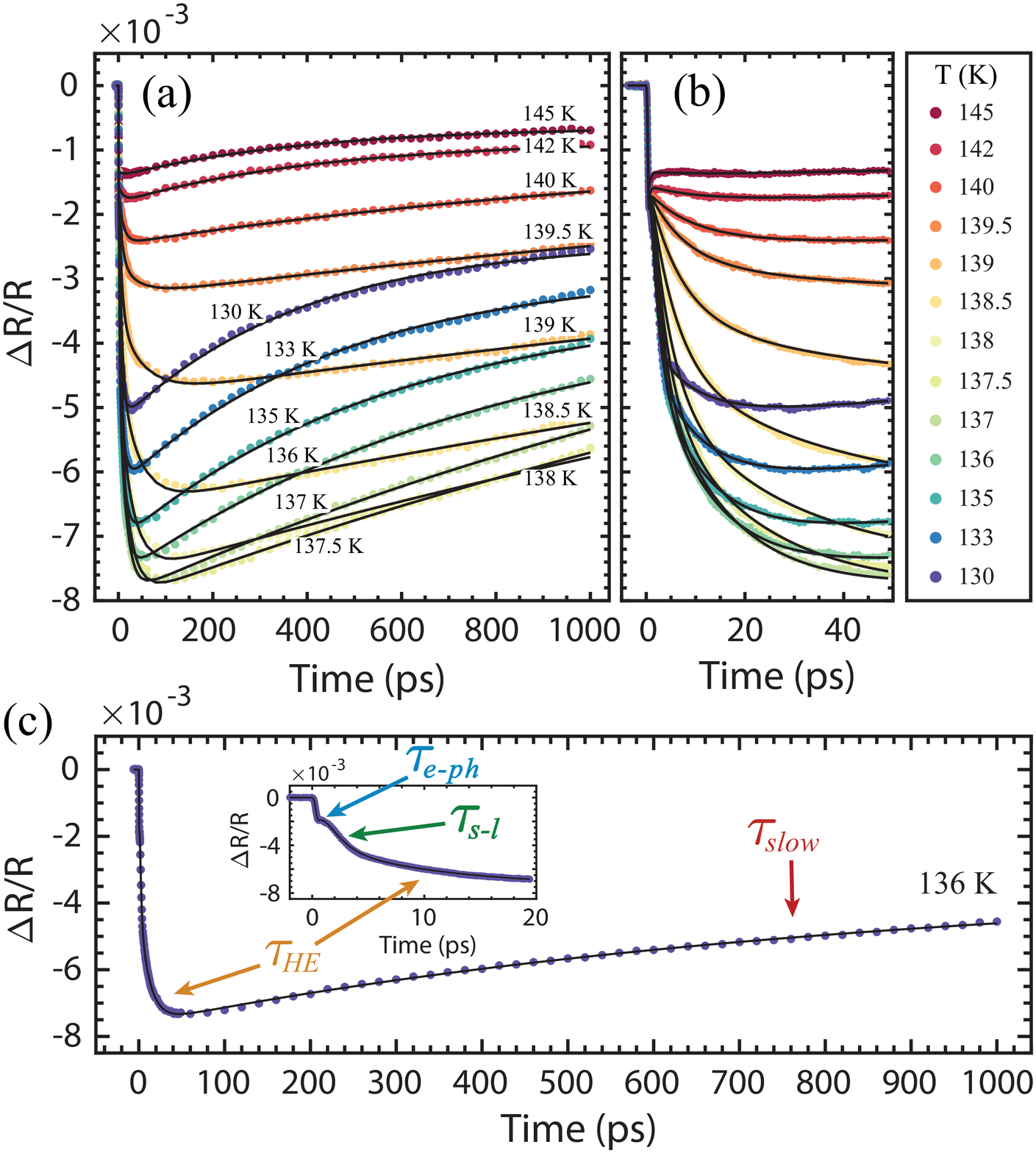}
  \caption[Photoinduced reflectivity traces near the critical temperature for long (a) and short (b) pump-probe delays.]
  {Photoinduced reflectivity traces near the critical temperature for long (a) and short (b) pump-probe delays with considerably smaller temperature steps in comparison to Fig. \ref{fig:LVO2_drrAll}. Black curves are multi-exponential fits to the data. (c) Pump-probe scan at 136 K, showing the four separate components of the response, color-coded relative to Eq. \ref{eq:LVO1_dRR}. \index{LVO3_drrTc}}
  \label{fig:LVO3_drrTc}
\end{figure}

To better understand the changes occurring between the high and low temperature phase, $\mathit{\Delta}R/R$ in the vicinity of \textit{T${}_{C}$} is plotted in Fig. \ref{fig:LVO3_drrTc}(a) for long time delays, and in Fig. \ref{fig:LVO3_drrTc}(b) for short delays. The data in this region is highly reproducible and taken with particularly small temperature steps. The changes are dramatic; there is nearly an order of magnitude increase in the signal amplitude, occurring as the result of a secondary rise time emerging in the 10 -- 40 ps range. The slowing of dynamics is clearly evident as well, the recovery rate entirely flattening as the signal maxima at $T=137.5$ K is approached, and the large offset at 1 ns indicating the lack of recovery of spectral weight. It is at this temperature that LVO enters the orbital and spin ordered phase, and the drastic changes in the signal reflect this. Further, this is the temperature at which formation of Hubbard excitons becomes energetically favorable after photoexcitation, due to the interaction with the AFM spin and anti-ferro orbital background in the ordered phase \cite{Novelli2012a,Reul2012a}. Thus the photoinduced signal here begins, in part, to reflect exciton dynamics, evident in the emergence of a new rise time.

A quantitative analysis of the signal supports these claims and reveals further details of the response. The dynamics below \textit{T${}_{C}$} can be fit by a four-component exponential decay plus a constant offset of the form: 

\begin{multline}
\label{eq:LVO1_dRR} 
\mathrm{\Delta }R/R\left(t\right) = \color{MidnightBlue}A_{e-ph}e^{-t/{\tau }_{e-ph}} \color{black}+ \color{OliveGreen}{A_{s-l}e^{-t/{\tau }_{s-l}}} \color{black}+ \\
\color{Bittersweet}A_{HE}e^{-t/{\tau }_{HE}} \color{black}+ \color{red}A_{slow}e^{-t/{\tau }_{slow}} \color{black}+ C, 
\end{multline}

\noindent The fits using this equation are shown as the black lines in Fig. \ref{fig:LVO2_drrAll} and \ref{fig:LVO3_drrTc}. An additional error function term (not included in Eq. \ref{eq:LVO1_dRR}) models the initial step-like rise in reflectivity. In the high temperature phase $T\ge 140\ K$, the 3${}^{rd}$ component labeled ${\tau }_{HE}$ vanishes, accurately fitting to only three exponentials. The full recovery process is largely dependent on the temperature, but in general there is a fast (${\tau }_{e-ph}<1$ ps) and slow recovery component (${\tau }_{slow}\approx 100$'s ps), along with an intermediate (${\tau }_{s-l}\le 10$ ps) and an emergent component (${\tau }_{HE}\approx 10-50$ ps), varying with temperature. These components are shown in Fig. \ref{fig:LVO3_drrTc}(c) for a representative $\Delta R/R$ signal at 136 K. 

As the measured time scales are well separated, we can attribute each time constant to a specific physical process. The initial delta-like increase in the magnitude of $\mathit{\Delta}R/R$, completed in less than 500 fs, can be attributed mainly to pump-induced photocarrier generation by intersite vanadium \textit{d-d} transitions along the \textit{c}-axis. This non-equilibrium distribution of carriers thermalizes via electron-electron (e-e) scattering within the initial rise time. At lower temperatures, in the ordered phase, this is accompanied by disruption of the orbital order, which similarly contributes to the sharp rise in $\Delta R/R$ by bringing the system further out of equilibrium \cite{Tomimoto2003,Yusupov2010b}. Following photoexcitation is an initial fast recovery, ${\tau }_{e-ph}$, which occurs in $\mathrm{\sim}$0.5 ps and is relatively temperature independent. This is attributed to electron-phonon (e-ph) relaxation, a process in which the hot photoexcited carriers thermalize with the lattice by coupling to optical and acoustic phonon modes. The timescale measured is consistent with results in other transition metal oxides, which measure e-ph relaxation on the order of $\mathrm{<}$ 1 ps \cite{Qi2012,Wall2009,MiyasakaNakamura2006}. The observed temperature independence of this time constant is consistent with expectations for electron-phonon thermalization.  

 A secondary rise time ${\tau }_{s-l}\le 10$ ps occurs subsequent to electron-phonon relaxation. Near room temperature, this component has a characteristic timescale of $\mathrm{\sim}$2 ps, which rises to $\mathrm{\sim}$10 ps near 180 K, where its amplitude crosses zero from negative to positive. There is no divergence in the time constant upon approaching the spin and orbital ordering temperature, though the amplitude does sharply peak at \textit{T${}_{C}$}. Instead, ${\tau }_{s-l}$ monotonically decreases to $\mathrm{\sim}$1 ps, and continues dropping to $\mathrm{\sim}$0.4 ps at the lowest temperatures. We ascribe this process to spin-lattice thermalization. Thus at a delay of $\mathrm{\sim}$ ${\tau }_{s-l}$ the electron, lattice, and spin subsystems are at the same temperature, between $\sim 1-5$ K above the initial pre-photoexcitation temperature in the vicinity of $T_C$ for the fluence we used. The observed timescales are consistent with measurements on other vanadates, such as YVO${}_{3}$ and GdVO${}_{3}$, where disruption of spin order occurs on the timescale of 2-4 ps \cite{MiyasakaNakamura2006,Mazurenko2008}.

\begin{figure*}
  \centering
  \includegraphics[width=0.9\textwidth]{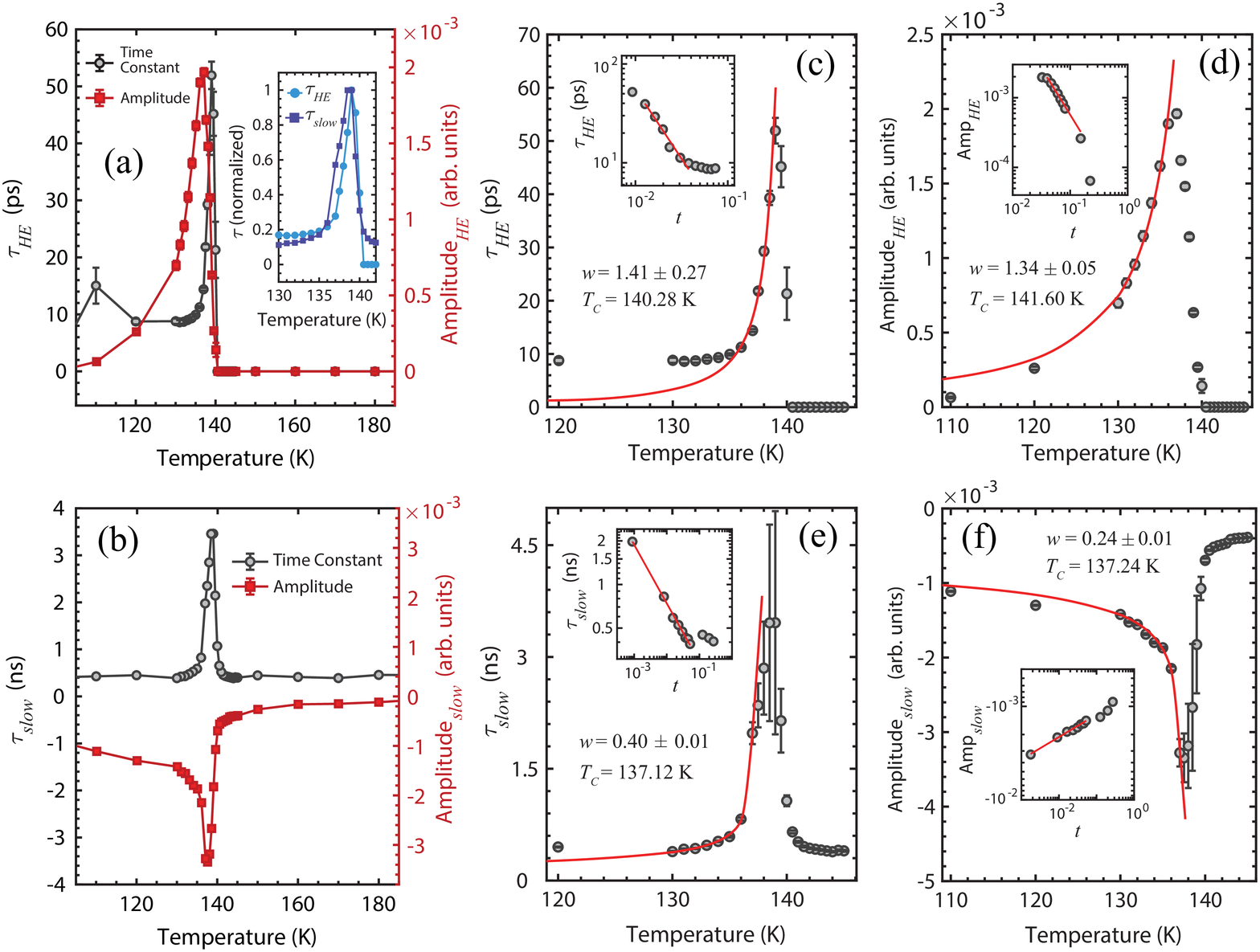}
  \caption[Time constants and amplitudes extracted from the exponential fits, for (a) the Hubbard exciton component $\tau_{HE}$ emerging at 140 K, and (b) the $\tau_{slow}$ component assigned to the recovery of spin and orbital order.]
  {Time constants and amplitudes extracted from the exponential fits, for (a) the Hubbard exciton component $\tau_{HE}$ emerging at 140 K, and (b) the $\tau_{slow}$ component assigned to the recovery of spin and orbital order. The lines are guides to the eye. The inset plots $\tau_{HE}$ and $\tau_{slow}$ on the same axis, normalized at 139 K for comparison. (c) and (d) depict the critical behavior and power-law fit (of the form given in Eq. \ref{eq:LVO2_powerLaw}, in red) for $\tau_{HE}$ and $A_{HE}$, respectively. The insets show the same data on a log-log scale with the reduced temperature. (e) and (f) present the power-law fits for the for $\tau_{slow}$ and $A_{slow}$, respectively. \index{LVO4_criticalFits}}
  \label{fig:LVO4_criticalFits}
\end{figure*}

 The third component of the signal ${\tau}_{HE}$ manifests as an additional rise on the order of 4 -- 50 ps emerging near the critical temperature at $T=140\ K$, shown in Fig. \ref{fig:LVO4_criticalFits}(a) as a function of temperature. We attribute this component to the transfer of spectral weight away from the Hubbard exciton peak in response to the photoinduced increased spin disorder and fluctuations (i.e. essentially a thermo-modulation signal from the small temperature increase). The value of ${\tau }_{HE}$ shows a sharp peak at \textit{T${}_{C}$}, while the amplitude peaks slightly below \textit{T${}_{C}$}. This component of the dynamics emerges only as G-OO and C-SO are established. Formation of the HE is governed by interaction of carriers with this antiferro-orbitally ordered and AFM spin ordered background. It follows that a timescale associated with the HE should emerge only below the ordering temperature, as observed. Keep in mind that ${\tau}_{HE}$ is not a direct measure of exciton formation timescales, but rather tracks electronic changes in the system that are intricately tied to the existence of the HE. On this timescale, the increased temperature of the spin and orbital subsystem significantly increases quantum fluctuations, reducing the binding affinity of HEs. This causes a transfer of spectral weight away from the HE peak in the optical conductivity to higher energies as the exciton population becomes smaller and more unstable, reducing the absorption at the probe energy and causing an additional decrease in $\mathit{\Delta}R/R$. We note that while the spin subsystem has thermalized with the other high-temperature degrees of freedom on ${\tau }_{s-l}$ timescales, spin disorder continues to propagate on longer timescales. This has been observed in YVO${}_{3}$ with time-resolved optical conductivity measurements, where spectral weight is transferred away from the HE peak on timescales between 50 -- 400 ps as local photoinduced spin perturbations diffuse \cite{Novelli2012a}. Propagation of orbital disorder also very likely plays a role, but an analysis of the critical behavior at $T_C$ in the proceeding section indicates that spin disorder is the primary contribution. At low temperatures spin and orbital order is locked in and fluctuations are reduced, minimizing the spectral weight transfer away from the HE peak and thus minimizing time constant and amplitude of the ${\tau }_{HE}$ component.

The final component ${\tau }_{slow}$ is an exponential recovery on the order of $\mathrm{\sim}$100 to $\mathrm{>}$1000 ps, with the extracted time constant and amplitude as a function of temperature shown in Fig. \ref{fig:LVO4_criticalFits}(b). At both high and low temperatures the recovery time is relatively constant (approximately $\mathrm{\sim}$400 ps). However, upon approaching $T_C$ there is a sharp divergence in both the time constant and amplitude, rising by nearly an order of magnitude. We attribute this final component to the recovery of the excitonic resonance spectral weight as fluctuations subside and spin/orbital order is re-established. The return of spectral weight to the probed HE peak corresponds to an increase in absorption, visible in the increase of $\mathit{\Delta}R/R$ as it begins to recover to its equilibrium value. While a significant offset in $\mathit{\Delta}R/R$ remains at the longest times measured, indicating that heat does not fully diffuse out of the lattice until $t\gg 1$ ns, the photoexcited spin and orbital subsystems do return to equilibrium within  $t\approx {\tau }_{slow}$. Above $T_C$, in the non-ordered phase, the amplitude of this component is nearly zero, only increasing to appreciable values as order and the HE develop, but photoinduced fluctuations must still recover and contribute to the recovery of $\Delta R/R$ at higher temperatures. The timescales we measure are highly consistent with optical pump-probe results on similar vanadates such as YVO${}_{3}$ and GdVO${}_{3}$, where the slow component assigned to spin relaxation is measured to be 300 -- 3000 ps \cite{Kim2018a}. Our data is also consistent with the 400+ ps recovery of spin disorder in YVO${}_{3}$ observed via spectral weight transfer between the HE and B peak \cite{Novelli2012a}. While the recovery of orbital order has been recorded closer to 50 ps in YVO${}_{3}$ and GdVO${}_{3}$ \cite{Kim2018a}, the close proximity of the spin and orbital ordering temperature in LVO and their highly coupled nature likely leads to additional frustration and longer orbital recovery times in our measurements. In summary, because the HE is highly dependent on the spin and orbital order in LVO and other vanadates, the recovery of order in these degrees of freedom re-establishes the excitonic resonance, transfers spectral weight back to the probed HE peak, and subsequently returns the $\Delta R/R$ signal to equilibrium.

\section{Analysis and Discussion}
\label{sec:LVOanalysisDiscussion}

\begin{figure*} 
  \centering
  \includegraphics[width=0.8\linewidth]{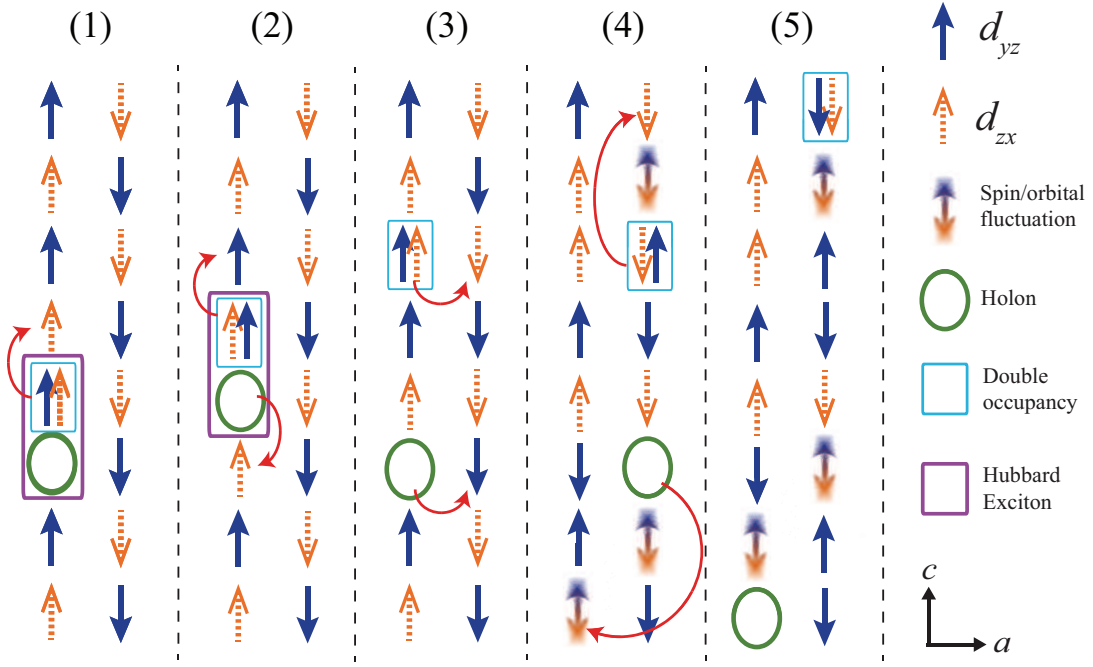}
  \caption[Exciton (purple square) dynamics near $T_C$ on an AFM spin and antiferro-orbitally ordered background.]
  {Exciton dynamics near $T_C$ on an AFM spin and antiferro-orbitally ordered background. Each arrow represents the spin and orbital occupation at a separate lattice site. (1) After photoexcitation, the holon and DO are bound as a Hubbard exciton. Hopping of the bound pair [(1) $\xrightarrow{}$ (2)] preserves spin and orbital order. Hopping of the individual holon or DO along the \textit{c}-axis [(2) $\xrightarrow{}$ (3)] results in orbital disorder, while hopping along the \textit{a} or \textit{b}-axis [(3) $\xrightarrow{}$ (4)] produces both spin and orbital disorder. At temperatures near $T_C$ and at times $t > \tau_{HE}$ after photoexcitation, spin and orbital fluctuations (depicted in (4)) disrupt the local order and the individual holon and DO are able to move further apart [(4) $\xrightarrow{}$ (5)]. This destabilizes the HE (5) and reduces spectral weight at the HE peak. \index{LVO6_HEdynamics}}
  \label{fig:LVO6_HEdynamics}
\end{figure*}

Fig. \ref{fig:LVO4_criticalFits} shows the temperature dependence of the amplitude and time constant of the Hubbard exciton (``HE") and slowly relaxing (``slow") components of the time dependent signal defined in Eq.~\ref{eq:LVO1_dRR}. The behavior of the two relaxation times is qualitatively similar. For temperatures greater than about $140$ K neither signal has significant amplitude; just below $140$ K a large amplitude peak appears, and then vanishes rapidly, decaying into background below about $135$ K. These measured recovery times ${\tau}_{HE}$ and ${\tau }_{slow}$ are, as described in the preceding section, associated with photoinduced spectral weight transfer away from and back to the HE peak, and are thus related to Hubbard exciton (HE) dynamics in the presence of a fluctuating many-body environment. Fig. \ref{fig:LVO6_HEdynamics} depicts HE dynamics in the spin and orbitally ordered phase. A holon and ``doublon'' are formed along the \textit{c}-axis after photoexcitation by the 1.6 eV pump. Motion of both quasiparticles as a bound pair, or exciton, preserves the spin and orbital order of the system [Fig. \ref{fig:LVO6_HEdynamics}(a)-(b)]. However, motion of either the bare holon or DO results in a trace of misaligned spins and orbitals. Motion along the \textit{c}-axis results in orbital disorder [Fig. \ref{fig:LVO6_HEdynamics}(b)-(c)], and motion along the \textit{a} or \textit{b}-axis produces both spin and orbital disorder [Fig. \ref{fig:LVO6_HEdynamics}(c)-(d)]. These quasiparticles will therefore seek to move together as a bound pair that leaves zero net disorder, which provides  the Hubbard exciton binding energy. Near $T_C$ and on ${\tau }_{HE}$ timescales spin and orbital fluctuations are large, disrupting the local order and relaxing the HE binding conditions. This allows the holon and DO to move further apart and destabilizes the HE [Fig. \ref{fig:LVO6_HEdynamics}(d)-(e)]. The result is a reduction of spectral weight at the probed HE peak both at temperatures near $T_C$ where spin/orbital fluctuations are inherently large, and at timescales on the order of ${\tau }_{HE}$ after photoexcitation when fluctuations are large due to pump-induced spin/orbital disorder and heating. Fluctuations must subside before the excitonic formation probability improves and spectral weight is transferred back to the HE peak, which occurs on ${\tau }_{slow}$ timescales after photoexcitation. As the sample is further cooled below $T_C$, fluctuations in the spin and orbital degrees of freedom are frozen as strong order is established, the HE is stabilized,  the optical conductivity at the HE peak greatly increases and the amplitude of the HE signal decreases. These qualitative considerations explain why the signal is maximal and long-lived near the transition temperature, and highlight the important role of spin/orbital disorder upon the carrier dynamics in LVO. 

We now consider the universal behavior near the phase transition. Experiment \cite{Bordet1993,Miyasaka2002} indicates that as the temperature is decreased below about  $140$ K both  structural/orbital order and  magnetic order occur. The theoretical literature indicates several possible scenarios, including simultaneous first order magnetic and structural/orbital transitions and sequential transitions in which a second order magnetic transition is closely followed by a weakly first order structural transition. Which scenario occurs is found to depend on fine details of interaction strength and material parameters  \cite{Khaliullin2001,Horsch2008}. The components of the time dependent signal associated with $\tau_{HE}$ and $\tau_{slow}$ are strongly temperature dependent, with dynamics becoming very slow near the ordering transition. Fig. \ref{fig:LVO4_criticalFits}(a) and (b) show the temperature dependence of the amplitude and time constant of the Hubbard exciton (``HE") and slowly relaxing (``slow") components of the time dependent signal defined in Eq.~\ref{eq:LVO1_dRR}. A dramatic increase of the characteristic time and of the amplitude is found as the temperature approaches the ordering temperature from below. While the behavior of both $\tau_{HE}$ and $\tau_{slow}$ are qualitatively similar in this temperature regime, a clear difference can be observed by normalizing both at 139 K and plotting on the same axis, shown in the inset of Fig. \ref{fig:LVO4_criticalFits}(a). A quantitative analysis can provide further insight.

One possible explanation of the dramatic temperature dependence is that the increase in time constant and amplitude arises from the critical slowing down associated with a second order magnetic phase transition. In this case one expects the characteristic time to scale as

\begin{equation} \label{eq:LVO2_powerLaw} 
\tau=\tau_0t^{w},    
\end{equation} 
where \textit{t} is the reduced temperature $t=\frac{|T-T_C|}{T_C}$, $w=z\nu$ is a dynamical exponent, and $\tau_0$ is the critical amplitude. $w$ is expected to exhibit universal scaling behavior and is therefore comparable to theoretical predictions. We fit the time constant and amplitude of the ${\tau }_{HE}$ and ${\tau }_{slow}$ components to a power-law of the form given in Eq. \ref{eq:LVO2_powerLaw}. These fits are shown as the red curves in Figs. \ref{fig:LVO4_criticalFits}(c)-(f). The value of $T_C$ is found by varying it within a small range to yield the best fit. The insets of these figures show the fit on a log-log scale as a function of the reduced temperature \textit{t}, where linearity indicates a good fit. Generally, we expect a close fit only in a small temperature region near $T_C$ where the theory is valid. We note also that data points very near to the peak signal are omitted from the analysis. In this highly critical region the experimental fitting procedure breaks down and produces divergent, unreliable fit parameters and errors, causing the data to deviate from expected power-law behavior. An additional explanation is provided by consideration of a weakly first-order transition, detailed in a proceeding section. Accordingly, while the precise values of the critical exponents and critical temperatures extracted from the analysis should be considered with some reservation, the results do indicate a clear difference in critical behavior between the ``HE" and ``slow" components.

Figs. \ref{fig:LVO4_criticalFits}(c) and (d) show the fits for ${\tau }_{HE}$ and $A_{HE}$, respectively. These yield critical exponents $w_{\tau -HE}=1.41\pm 0.27$ and $w_{A-HE}=1.34\pm 0.05$, and critical temperatures $T_{C-\tau -HE}=140.28$ K and $T_{C-A-HE}=141.60$ K. This yields an average critical temperature of $T_{C-HE}=140.94\pm 0.93$ K for the \textit{HE} component. The same analysis can be applied to the \textit{slow} component, and is shown in Figs. \ref{fig:LVO4_criticalFits}(e) and (f) for ${\tau }_{slow}$ and $A_{slow}$, respectively. These yield critical exponents $w_{\tau -slow}=0.40\pm 0.01$ and $w_{A-slow}=0.24\pm 0.01,$ and critical temperatures $T_{C-\tau -slow}=137.12$ K and $T_{C-A-slow}=137.24$ K. This yields an average critical temperature of $T_{C-slow}=137.18\pm 0.08$ K for the \textit{slow} component. 

The results suggest a difference in critical temperature between the ``HE" and ``slow" components, but given the range in which experimental data is omitted from the analysis at the signal peak we can make no definitive conclusions about differences in phase transition temperature. The critical exponents of the ${\tau }_{HE}$ and $A_{HE}$ component in Figs. \ref{fig:LVO4_criticalFits}(c) and (d) are not far from the $w = z\nu\approx 1.26$ expected for the 3D Ising model \cite{Pelissetto2002,Hohenberg1977a}, which describes the anisotropic \textit{C}-type AFM spin order in LVO. The critical exponents measured from the $\tau_{slow}$ components in Figs. \ref{fig:LVO4_criticalFits}(e) and (f) are substantially lower, and do not closely match any other magnetic models \cite{Campostrini2002, Hohenberg1977a, Onsager1944, Mazenko1981a}. The critical behavior of the ${\tau }_{HE}$ component near $T_C$ appears magnetic in nature, nearly matching that of the 3D Ising universality class, but the ${\tau }_{slow}$ component does not belong to any predicted universality class for magnetic order. These results are consistent with our assignments of ${\tau }_{HE}$ and ${\tau }_{slow}$. ${\tau }_{HE}$ represents the diffusion of spin disorder, a phenomenon described by a magnetic order parameter that obeys universal scaling laws. On the other hand, the departure of ${\tau }_{slow}$ from universal scaling behavior implies that this timescale reflects more than just the spin degree of freedom. It is a measure of the recovery of both spin \textit{and} orbital order, and therefore need not belong to a particular universality class. However, the power-law dependence of ${\tau }_{slow}$ does suggest that orbital recovery dynamics may be modeled by an order parameter obeying universal scaling behavior with an associated critical exponent. To our knowledge, such a theory does not exist to model orbital critical dynamics.

\begin{figure}[t] 
  \centering
  \includegraphics[width=\linewidth]{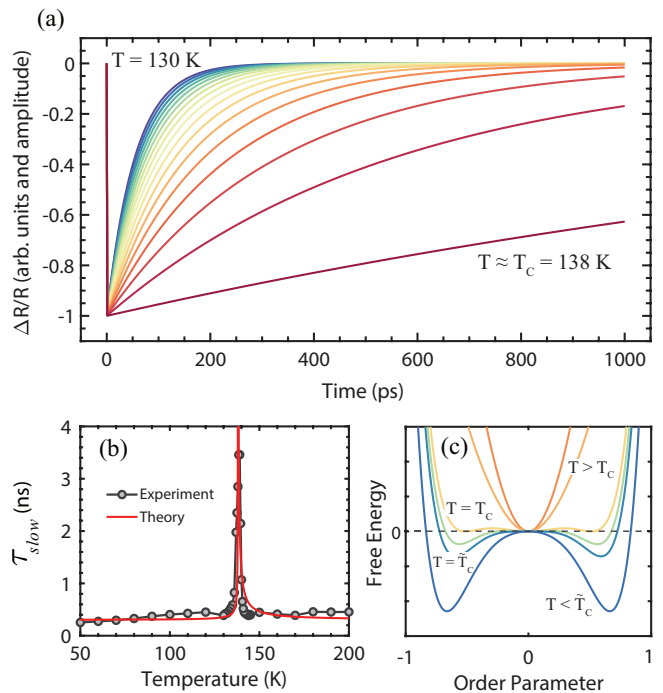}
  \caption[Time-dependent Ginzburg-Landau simulations for a first order phase transition, as a function of temperature near $T_C$.]
  {(a) Time-dependent Ginzburg-Landau simulations for a generic first order phase transition, as a function of temperature near $T_C$ ($\alpha_0=0.25$, $\beta=-0.35$, $\gamma=1$, $T_C=138 K$ ). The results qualitatively match the experimental data, with a similar slowing down of dynamics approaching $T_C$. (b) The simulated $\tau_{slow}$ is plotted in red alongside the experimental values, showing a remarkable correspondence. (c) The free energy given by Eq. \eqref{eq:freeEnergy} as a function of the order parameter $m$. Note that the free energy never completely flattens at $T_C$, instead retaining three discrete minima. \index{LVO5_simulation}}
  \label{fig:LVO5_simulation}
\end{figure}

In the above analysis we have assumed a phase transition that is second-order in nature and which exhibits rounding at small values of the reduced temperature, causing points near the signal peak to deviate from power-law scaling and adding uncertainty to the conclusions. One possibility is that inhomogeneous broadening associated with a distribution of phase transition temperatures complicates the analysis at very small values of the reduced temperature. Another possibility, which we explore in detail here, is that the transition is in fact weakly first order \cite{Hohenberg1977a}, as indicated by  heat capacity, magnetization, and neutron diffraction measurements on LVO single crystals \cite{Miyasaka2003,Tung2008}. We consider the free energy

\begin{equation}
   \label{eq:freeEnergy}
   f(T)=f_0\left(T\right)+{\alpha }_0\left(T-\tilde T_C\right)m^2+\frac{1}{2}\beta m^4+\frac{1}{3}\gamma m^6
   \end{equation}
where \textit{m} is the order parameter, $\gamma>0$ and $\beta<0$ for a first order transition and $>0$ for a second order one.  $f_0\left(T\right)$ describes the temperature dependence of the high temperature phase near the phase transition. for $\beta<0$ the model has a first order phase transition at temperature $T_C=\tilde T_C+\frac{3\beta^2}{16\alpha_0\gamma}$, and $\tilde T_C$ is the lower spinodal temperature below which the disordered phase becomes linearly unstable. 

The dynamics is then described by the relaxational (``model A") equation
\begin{equation}
\frac{1}{D}\frac{\partial f}{\partial t}=-\frac{\partial f}{\partial m}
\end{equation}
with $D$ a diffusion constant. Within the linear response regime the relaxation dynamics back to equilibrium $m=m_0$ follow an exponential form with 
\begin{equation}
m(t)-m_0\sim e^{-t/\tau}
\end{equation}
and characteristic recovery time
\begin{equation}
\tau=\begin{cases} \frac{1}{D}\frac{1}{2\alpha_0 (T - \tilde T_C) } &T>T_C\\ \frac{1}{D}\frac{1}{ 8 \alpha_0(\tilde  T_C - T) + \frac{2\beta}{\gamma}\left(\beta-\sqrt{\beta^2+4\alpha_0\gamma(\tilde T_C-T)}\right) } &T<T_C,\\\end{cases}
\end{equation}
In the limit $\beta\to 0$ the transition turns from a first to a second order one and critical slowing down as typically present in second order phase transition is found in the limit $T\to \tilde  T_C$. At nonzero but small $\beta$ (weak first order phase transition) the above mentioned slowing down persists as $T$ approaches $\tilde  T_C$ from above up to the point where the temperature $T$ approaches $T_C=\tilde  T_C+\frac{3\beta^2}{16\alpha_0\gamma}$, at which point the diverging relaxation time is cut off and the phase with finite order parameter $m_0$ becomes the stable one. The maximal relaxation rate approaching $T_C$ from above [below] is $8\gamma/(3 \beta^2)$ [$2\gamma/(3 \beta^2)$] diverging as $\beta^2$ for small $\beta$. 

As shown in Fig. \ref{fig:LVO5_simulation}(a), we find that the simulation using a small, fixed value of $\beta$ yields dynamics which, qualitatively, are remarkably similar to the experimental pump-probe results: a prominent slowing down and flattening in the recovery rate as $T_C$ is approached. Fig. \ref{fig:LVO5_simulation}(b) shows the simulation-derived time constant $\tau_{slow}$, plotted versus temperature along with the experimental data, where we associate $\Delta R/R\sim m-m_0$. We have added a small offset to the theoretical $\tau$ to reflect that the Ginzburg-Landau theory is a first order expansion around $T=\tilde T_C$, and at larger deviations from $T\approx T_C$ the prefactor of $m^2$ in $f$ should not increase/decrease without bound (which is the reason for $\tau\to 0$ at temperatures far away from $T_C$). The modeled results show a similar divergence in ${\tau }_{slow}$ at $T_C$, capturing the essential nature of the critical slowing down. This model can also be compared to the $\tau_{HE}$ component, which fits the experimental data similarly but cannot provide further insight or distinctions with out including the microscopic details of the LVO system. The success of this model indicates that the transition may not be truly of second-order but instead weakly first order (small $\beta$), associated with pseudo-critical dynamics which may not necessarily obey universal scaling laws \cite{Ikeda1979}. Recent work on charge and magnetically ordered perovskite La${}_{1/3}$Sr${}_{2/3}$FeO${}_{3}$ films have shown similar pseudo-critical behavior near the magnetic phase transition \cite{Zhu2018}. 

A study of the free energy provides further insight into the dynamics and the origin of the observed critical-like slowing down. Fig. \ref{fig:LVO5_simulation}(c) depicts the simulated free energy given by Eq. \eqref{eq:freeEnergy} as a function of the order parameter $m$ for various temperatures in the vicinity of $T_C$ and $\tilde  T_C$. Generally, as a phase transition is approached, the free energy landscape widens and bifurcates from a single potential minimum to two, separated by some non-zero value of the order parameter. In doing so, any perturbation of the order parameter will take longer to reach equilibrium in this shallower potential at \textit{T${}_{C}$}, thus slowing down the dynamics. We note that in the weakly first order limit the free energy never completely flattens at $T_C$, instead retaining three unique minima. These minima, however, become exceedingly shallow at smaller $\beta$. The linear fluence regime of the experiment suggests that we remain in the local minimum at $m = 0$, though at temperatures near $T_C$ the potential barriers become small and the system may be more easily pushed into and out of this local region. The defining consequence of having a weakly first order instead of a second order transition is thus that the slowing down persists only up to a cutoff which scales as $1/\beta^2$. At the point of the cutoff the order parameter jumps to a finite value as the minima at $m\neq 0$ lower in energy. For each temperature these minima have a finite curvature (finite relaxation time $\tau$) which however can become very shallow (with very large $\tau$) at small $\beta$ (with the $\beta\to 0$ limit recovering the unhampered slowing down known from a second order phase transition). In this manner, the weakly first order nature of the phase transition implies that there is a pseudo-divergence over a shorter temperature range, and that the power-law scaling expected of a second-order phase transition will fall off very near $T_C$, as we observe. 

\section{Conclusion}
\label{sec:LVOconclusion}

In summary, we have measured the time-resolved photoexcited dynamics of LaVO$_{3}$ thin films using ultrafast pump probe spectroscopy. The 1.6 eV pump and probe wavelength corresponds to an excitonic resonance above the Mott-Hubbard gap. As the Hubbard exciton is highly dependent on the spin and orbital order of the system, measurements at this photon energy serve as a sensitive probe of order parameter dynamics. The dynamics after photoexcitation proceed as follows: (1) direct photoexcited carrier generation, followed by electron-electron thermalization within 500 fs; (2) thermalization of hot electrons with the lattice through electron-phonon coupling, occurring in $\mathrm{\sim}$0.5 ps; (3) perturbation of the spin order and out-of-equilibrium thermalization of the spin, lattice and electron subsystems through spin-lattice coupling, occurring on the order of 2 -- 10 ps; (4) transfer of spectral weight away from the HE peak in the optical conductivity as spin disorder diffuses and the excitonic resonance decreases in $\sim$4 -- 50 ps; (5) and recovery of the HE spectral weight as spin and orbital order recover and photoinduced fluctuations subside within $\mathrm{\sim}$300 -- 3500 ps.  The dynamic HE spectral weight transfer reveals anomalous slowing down of the exciton, spin, and orbital dynamics at the structural, spin, and orbital ordering temperature $T_C\cong 140$ K. All display pseudo-critical behavior and critical slowing down, reminiscent of a second-order phase transition, despite the (weakly) first-order nature of the transition. The results indicate that the Hubbard exciton, and the carrier dynamics in general, are extremely sensitive to the coupled spin and orbital order in the system. As such, fluctuations in these degrees of freedom contribute to the critical slowing down at $T_C$.

This work provides a path for further investigating Mott-Hubbard exciton dynamics in transition metal compounds with highly correlated spin and orbital order. The observation of non-universal critical behavior in the coupled spin and orbital order recovery component suggests that orbital dynamics might potentially be described by a new dynamical scaling theory with its own unique critical exponent. The specific relation of the individual spin and orbital degrees of freedom to the exciton binding energy also merits further study. High sensitivity, low fluence measurements on related Mott-Hubbard insulators with well separated phase transition temperatures, such as YVO${}_{3}$ and YTiO${}_{3}$, should be performed to check for similar critical behavior. Studying Hubbard exciton physics in these highly correlated systems will allow us to better understand the complex materials where multiple intertwined degrees of freedom interact to produce pairing mechanisms and interesting emergent behavior.

\section{Acknowledgements}
\label{sec:LVOacknowledgements}

This work is primarily supported by the US Department of Energy (U. S. DOE), Office of Basic Energy Sciences (BES), under Grant No. DE-SC00012375. DMK  acknowledges funding by the Deutsche Forschungsgemeinschaft (DFG, German Research Foundation) under Germany's Excellence Strategy  -
Cluster  of  Excellence  Matter  and  Light  for Quantum Computing (ML4Q) EXC 2004/1 - 390534769. We acknowledge support by the Max Planck Institute - New York City Center for Non-Equilibrium Quantum Phenomena.

\end{document}